# Replay and compositional computation

Zeb Kurth-Nelson[1,2], Timothy Behrens[3,4], Greg Wayne[1], Kevin Miller[1,5],
Lennart Luettgau[2], Ray Dolan[2,3], Yunzhe Liu[6,7], Philipp Schwartenbeck[8,9]

[1]DeepMind, London, UK
[2]Max Planck UCL Centre for Computational Psychiatry and Ageing Research, London, UK
[3]Wellcome Centre for Human Neuroimaging, University College London, London, UK
[4]Wellcome Centre for Integrative Neuroimaging, University of Oxford, Oxford, UK
[5]Institute of Ophthalmology, University College London, London, UK
[6]State Key Laboratory of Cognitive Neuroscience and Learning, IDG/McGovern Institute for Brain Research, Beijing Normal University, Beijing, China
[7]Chinese Institute for Brain Research, Beijing, China
[8]Max Planck Institute for Biological Cybernetics, Tubingen, Germany
[9]University of Tubingen, Tubingen, Germany

Replay in the brain has been viewed as rehearsal, or, more recently, as sampling from a transition model. Here, we propose a new hypothesis: that replay is able to implement a form of compositional computation where entities are assembled into relationally-bound structures to derive qualitatively new knowledge. This idea builds on recent advances in neuroscience which indicate that the hippocampus flexibly binds objects to generalizable roles and that replay strings these role-bound objects into compound statements. We suggest experiments to test our hypothesis, and we end by noting the implications for AI systems which lack the human ability to radically generalize past experience to solve new problems.

Replay in the brain was discovered in the context of spatial experiments on rodents. While the animal is moving, hippocampal neurons encode its current location in space. But when the animal rests or pauses, researchers noticed that the same neurons sometimes spontaneously sweep through a sequence of firing patterns which repeat a path the animal recently traversed through space, at an accelerated rate[83,99,128]. The phenomenon was fittingly termed 'replay', and is a major component of hippocampal function, comprising a large fraction of all spikes at rest[18,20]. It was hypothesized that hippocampus rapidly stores new experiences, and that replay is rehearsal to transfer this knowledge into a stabler form in cortex[92,158], a process called consolidation.

However, over the past 20 years, our understanding of replay in the brain has grown. Although early experiments found replay sequences that directly repeat past experience, it gradually became clear that this is a special case of a much broader phenomenon. For example, Gupta et al[46] measured replay sequences in rats who had experience running in a T-shaped maze. The animals always ran north up the central corridor and turned left or right at the T-junction. At rest, replay sequences traversed the entire width of the east-west corridor, stitching together segments of space that had only been experienced separately. Replay even synthesizes entirely novel sequences. In another experiment[115], rats alternated between searching an open 2D environment for a randomly placed reward, and returning 'home' (a fixed location) for another reward. The home location changed each day, creating many possible combinations of random location and home location. Spontaneous replay sequences formed trajectories ahead of the animal, predicting its future path, and this was true even when that particular path had never been physically traversed before. Similarly, when barriers are moved dynamically in a 2D environment, replay sequences quickly adapt to play trajectories that route around the new barrier locations[156]. When the only behavioral task is unstructured random foraging in an open environment, replays at rest play out diffusion-like trajectories which were never followed by the animal[135]. Most remarkably, replays can traverse areas of space that have never been visited. When rats are given full view but only partial access to a maze, and food is dropped into the inaccessible part, replay sequences spontaneously represent trajectories into the inaccessible part that has never physically been visited[102].

In light of an accumulating body of such data, the predominant view in neuroscience is now that sequences are not constrained to simply repeat past experience, but are informed by an internal model of the world[30,36,105,114,118,159]. The use of an internal model opens the door for replay to describe hypothetical scenarios based on the causal structure of the world[9,113,140]. This model could be used for online control of behavior ('planning')[62,91,105,115] and for offline simulation to train a policy or value function (like 'Dyna')[61,90,139].

## A new conjecture

The goal of this paper is to propose another update to how we think about replay. Our proposal is that replay in the brain instantiates a form of compositional computation. We will argue that a replay sequence constitutes a set of entities strung together into a compound. Each entity is transiently bound to a representation of its role in the compound, which specifies how that element functions as a part of the whole (Figure 1a,b). Thus, the replay sequence as a whole describes a structure whose meaning is an interaction of the parts and their relationships. Composing entities in this way allows replay to derive new knowledge. For example, if I already know that my laptop contains a CPU and that the CPU draws 30 watts, then a replay sequence could derive the implication that the laptop draws at least 30 watts. This is a speculative hypothesis, but it is a natural extension of recent neuroscientific discoveries about replay, which we will review. We will also offer experiments to test several predictions of the idea.

The replay we envisage is built on two types of composability. The first type is a separation between entity and role (or, semantics and syntax): new entities can be bound to existing roles and vice versa (Figure 1c). The second type is that there are many ways to arrange role-bound entities into compounds (sequences), creating a potentially infinite space of compounds from a smaller number of elements (Figure 1d). We will use the idea of these two types of composability to organize the following sections.

Our view agrees with the notion that replay draws on an internal model. However, in previous model-based accounts of replay, a replay event is a 'rollout': a sequence of states predicted to occur sequentially in the world, with the model defining the transition probabilities between successive states. Our hypothesis relaxes this constraint, so that items in a replay event need not be successive states, but instead may be a set of entities with arbitrary relationships to one another. Our hypothesis also relaxes the constraint that replays are used for online control of behavior or for updating a policy or value function. Instead, we suggest that replays can be used to derive new knowledge about the world by reasoning about the implications of combining existing pieces of knowledge. To forestall misunderstanding, our conjecture is not that all replay is best conceived as compositional computation. In simpler settings, replay does appear to recapitulate experience or to perform rollouts. Rather, we suggest that the machinery of replay – which likely evolved to solve simpler problems – gives rise to compositional computation when coupled with learning mechanisms and rich environments, most dramatically in humans.

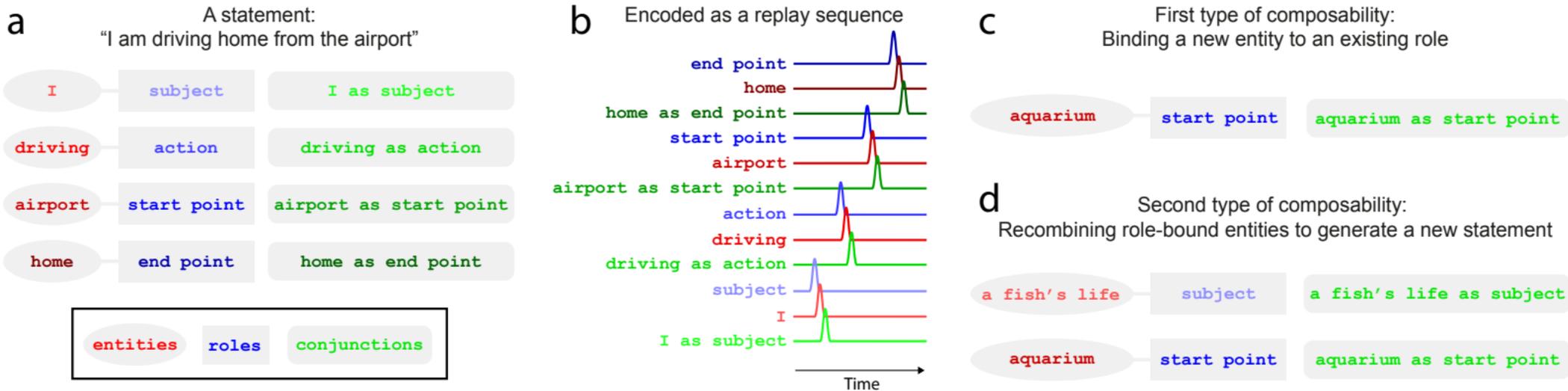

**Figure 1: Two types of composability in replay. a,** A structured compound, such as the concept 'I am driving home from the airport', can be described as a set of entities and the role each entity plays in the compound. Each entity, like driving, is bound to a role, like action. The compound has a meaning that derives from the interaction of its parts but is not contained in any of the parts. **b,** In our hypothesis, a replay sequence constitutes a set of entities, each tagged with the role of that entity, and can describe complex concepts. **c,** One type of composability is that roles are independent of entities and are free to attach to new entities. **d,** Another type of composability is that role-bound entities can be combined in many different ways to form new compounds.

## Binding entities to generalizable roles in hippocampus

The first type of composability is binding new entities (role-fillers) to existing roles or vice versa[9,55,129]. In the example of Figure 1, the entity *airport* takes the role of *start point*, as part of a compound meaning 'I'm driving home from the airport'. The same role attached to a new entity, *aquarium*, would immediately have a different but interpretable meaning ('aquarium as start point') and could be used to make new compounds.

Our perspective is that hippocampus has a general-purpose ability to bind entities to roles. This is an unproven idea, but experimental data from humans and animals make it an exciting possibility. We discuss these data in terms of three potential levels of generality of roles. At the first level, roles are constrained to be spatial (where something is relative to other things). At the second level, roles can be nonspatial but adhere to the Euclidean geometry of space (for example, a coordinate in the 1D space of auditory pitch). At the third level, roles can be nonspatial and non-Euclidean, potentially including arbitrary roles like *start point* or *verb*.

The large hippocampal literature in animals has studied primarily spatial roles, our first level of generality. Hippocampus receives converging input from the brain's 'what' stream, through lateral entorhinal cortex, and 'where' stream, through medial entorhinal cortex[40,67,71,86]. The 'what' stream carries representations that discriminate between different objects or sensory details but are invariant to how they are arranged in space. The 'where' stream carries an abstract representation of space itself, providing a coordinate system that describes relative positions. These two streams combine in hippocampus to form conjunctive codes, where the abstraction of space is joined with sensory specifics to code for a particular place or memory[9,72,154] (Figure 2a). For example, medial entorhinal cortex contains object-vector cells (OVCs), which fire when the animal is at a particular location relative to any object in the environment[111] (Figure 2b). These cells code for a relational role, invariant to sensory specifics. Meanwhile, landmark cells in hippocampus also fire at a particular location relative to an object, but restrict their firing to certain objects[27]. These cells code for a conjunction of role and sensory filler. In sum, the rodent spatial literature offers extensive evidence that hippocampal representations conjoin abstract roles with sensory specific to describe role-bound entities.

Newer experiments, in both humans and non-human animals, have started to broaden our understanding of the 'where' pathway beyond physical space. Experimental data from the past decade[3,13,23,69,141], aligning with older theories[22,143], suggest that the 'where' pathway carries information about nonspatial relational roles, ranging from social relationships to auditory pitch to configural stimulus spaces. Correspondingly, hippocampus has conjunctive representations that go beyond space[3,85,160]. So far, experiments with nonspatial relations have mostly been limited to the second level of generality – Euclidean topologies such as 2D social relationships and 1D auditory pitch.

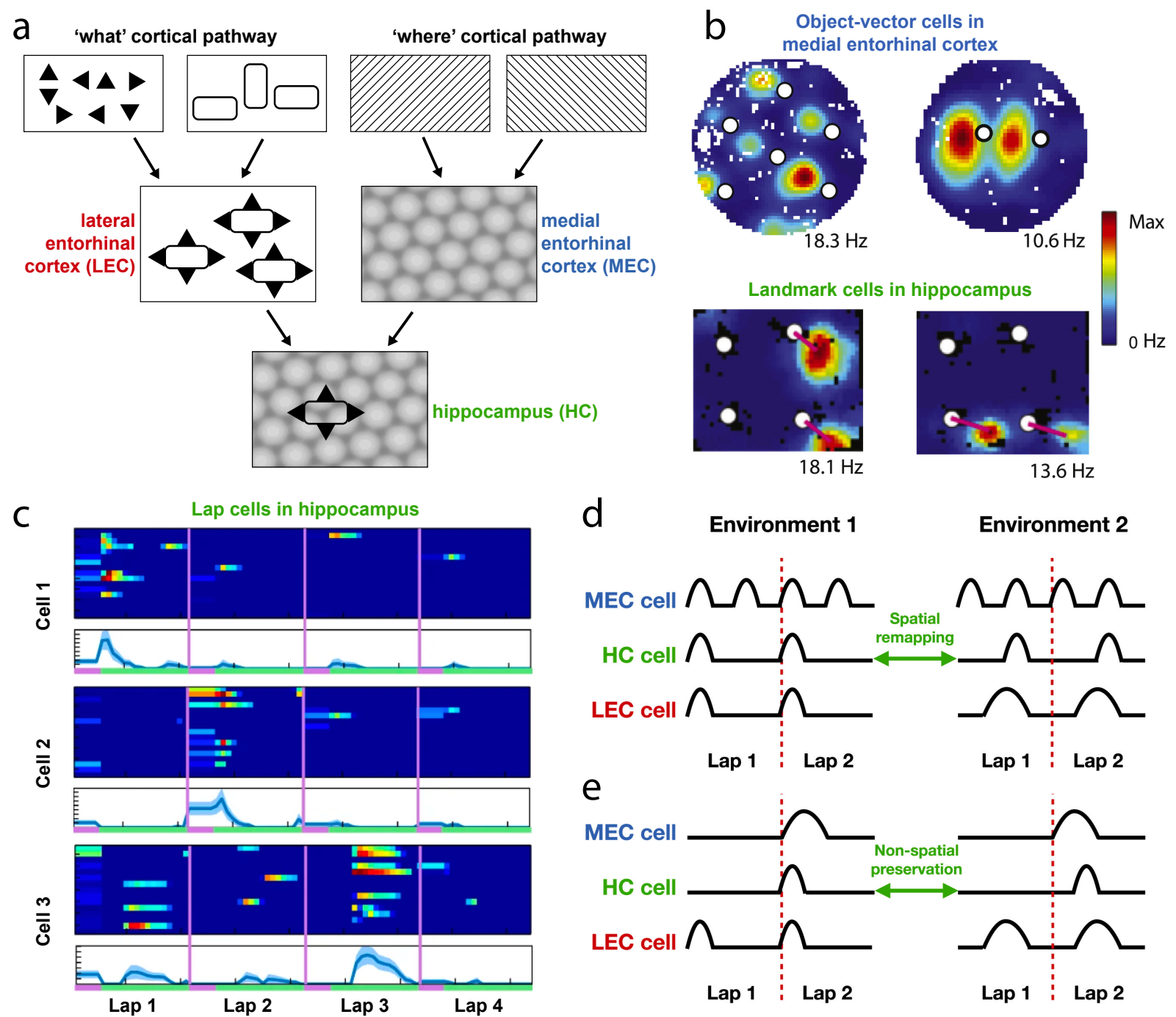

**Figure 2: Binding of entities and roles in hippocampus. a,** A highly schematized depiction of two processing pathways in cortex, adapted from Manns and Eichenbaum[86]. The 'what' pathway, including posterior parietal cortex and perirhinal cortex, extracts and transmits information about the identities and semantics of perceived entities (shown as geometric objects). The 'where' pathway, including inferior temporal cortex and parahippocampal cortex, extracts and transmits information about relational roles of entities (shown as a repeating pattern representing an abstract code for space). After these pathways reach cells in lateral and medial entorhinal cortex, respectively, they converge on cells in hippocampus to produce conjunctive representations of 'entity in role', such as an episodic memory or a location in a particular environment. **b,** Object-vector cells, in medial entorhinal cortex, fire at a particular location relative to each object in the environment[111]. Landmark cells, in hippocampus, show a similar pattern but are selective for particular objects[27]. **c,** In the experiment of Sun et al[136], mice ran four identical laps around a track before getting reward. Some hippocampal cells ('lap cells'; three examples shown) were tuned not only to place but also to lap index, firing selectively on particular laps. x-axis within each lap is spatial location of the animal. Color axis shows firing rate at that location, on that lap. Each row within each cell is a trial. **d,** Each mouse performed the four-lap task in two different environments. Consistent with many prior experiments, hippocampal place cells remapped to new spatial locations between environments. Remapping can be explained as the same spatial role codes (MEC) being conjoined with by new sensory role-filler codes (LEC) in the new environment. Again, x-axis within each lap corresponds to spatial location. y-axis represents firing rate. **e,** Lap cells did not remap to a different lap index between environments, which is explained by discrete role codes tuned to a particular lap[154].

However, there are hints that roles may even go beyond Euclidean to the third level of generality. One point of supporting data is that non-Euclidean role representations appear in replay in humans[81], which we discuss in more detail in the next section. Another piece of evidence comes from the representation of discrete states in rodent hippocampus. In a task where a mouse had to complete four laps of a loop before reward became available[136], hippocampal cells responded (as usual) at selective places in the lap. Some of these place cells were also 'lap cells': they responded on a particular lap (Figure 2c). When the animal was asked to do the same task on a maze with different sensory cues, place cells that were not lap cells exhibited spatial remapping as usual[38] (Figure 2d). Lap cells also changed their spatial tunings, but crucially they always fired on the same lap in both mazes (Figure 2e). This is exactly what is predicted by a system that has roles for spatial location and lap number, and fills them with the sensory particularities of each individual maze[154]. The different sensory input along each maze shifts the conjunctive cells in space, but this mechanism can never shift a cell between laps because each lap has the identical sensory input. Indeed, this same mechanism also explains open-field remapping experiments, where place cells preferentially remap to different peaks of the same grid cells. By remapping, but staying at the same grid phase, a single place cell preserves its role input across different situations. Alternate accounts of these data, such as predicting reward or representing total distance traveled, are difficult to reconcile with the fact that they shift at random between different spatial locations (and different distances from the reward) in the two mazes.

These hints of non-Euclidean role representations are limited to coding for an index within a sequence[81,136] or for a binary category[81]. But because roles can be learned from data[9,133,154], there is potentially no limit to their richness and diversity. Future experiments are needed to probe a broader spectrum of roles, like *verb* (see section Experimental Predictions). Even if rodents represent some kinds of non-Euclidean roles, their repertoire is likely narrower than humans. But to the extent that the 'where' pathway does carry information about general-purpose roles, conjunctive codes in hippocampus may act as general-purpose bindings of entities to roles.

## Replay as compositional computation

The central hypothesis of this paper is that replay sequences are made up of hippocampal representations of role-bound entities (first-level composition) – and by chaining these together into structures (second-level composition), sequences can express a huge variety of new compound meanings. Sequencing is a natural way to link entities because it interferes minimally with the spatial activity patterns that encode individual items. In other words, activating items in sequence keeps separate items separate. This is similar to how, in language, we put words in sequence instead of superimposing them. Continuing with the example of Figure 1, combining *aquarium/start point* with *a fish's life/subject* makes a statement that is new yet interpretable.

Our hypothesis builds on the idea of hippocampus as a sequence generator[21], where its function is to construct generic content-free sequences into which any content can be slotted. We add two things to this model. First, sequences don't have to be played in the order of experience; replay can assemble any set of things in any order. Second, each item in a sequence is bound to a representation of its role in the compound, allowing arbitrary compositional concepts to be constructed by putting together elements in complex ways.

Compositional computation leverages the fact that the world itself is made up of more-or-less separate entities whose recombination in different ways is useful. Reassorting knowledge into meaningful new compounds brings the potential for strong generalization[8,19,41,149,165] and may underlie the flexibility of human imagination[16,37,50].

**Separate entities in replay**

The power of compositional systems comes from treating the world as made up of separate entities which can be recombined (Figure 3a). Reasoning about relationships between separate things is a potent form of abstraction: discarding everything except the relevant entities and relations[8,100,146]. For example, an animal might receive a reward due to an action it took an hour ago. This is a difficult credit assignment problem[56] unless the animal abstracts away the continuum of sensory data between the action and the outcome.

It is therefore interesting to view a replay sequence as a set of separate representations composed together into a unit. In nonspatial tasks with inherently discrete structure, replay sequences jump abruptly between representations of the objects[73,81,123], as we will describe in detail in the following sections. More surprisingly, replay also hops between discrete positions in spatial tasks[116] (Figure 3b). These hops are aligned with the gamma oscillation. At each peak in gamma, replay dwells at a stationary point in space, and at troughs, the representation jumps to a new point. Pfeiffer and Foster[116] suggest that peaks in gamma focus representation on a unit of information, and troughs allow transition to a different unit. This idea follows from the large literature on pattern separation and pattern completion within hippocampal circuitry[76,144,163] as a mechanism for sharpening the differences between separate items. The Pfeiffer and Foster data are also consistent with longstanding theories that the gamma oscillation organizes neuronal spiking into discrete slots which correspond to cognitively relevant parcellations of information[19,49,79,80].

Under this view, it is plausible that both sharp wave ripple sequences and theta sequences form compositional compounds. Most of the replay experiments we will discuss in this paper have investigated ripple-nested sequences, but a theta sequence also constitutes a set of conjunctive hippocampal representations. Theta sequences are associated with active behavior and may therefore perform compositional sampling for active control[62,66,157], a topic we will return to later.

**Replay beyond space**

For replay to implement general-purpose compositional computation, an obvious requirement is that it operates beyond the spatial domain measured in rodent replay experiments. Methodological advances have made it possible to detect replay in humans using

a 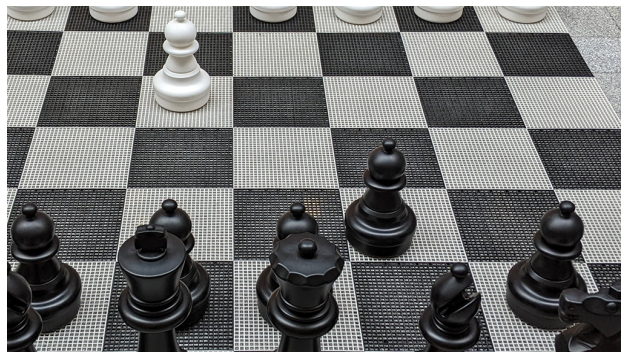 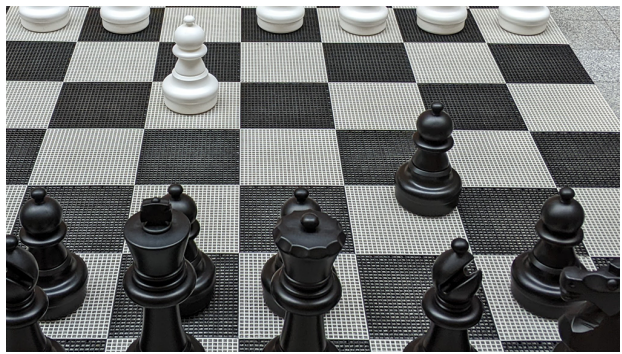 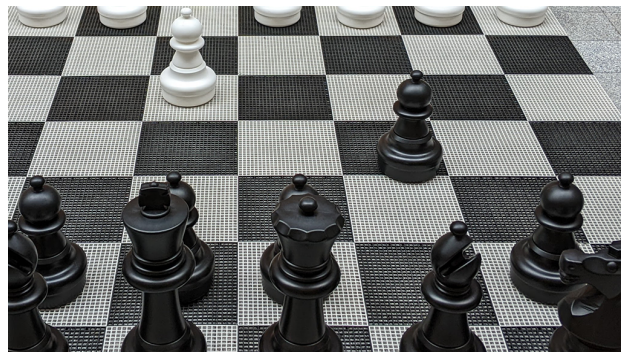 b 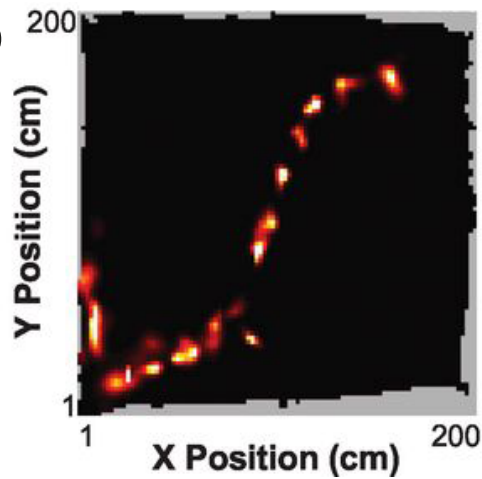

Description in terms of entities: pawn on e4, pawn on c6 | pawn on e4, pawn on c6 | pawn on e4, pawn on c5

**Figure 3: Discreteness in replay. a,** Compositional systems divide the world into entities that can be recombined, like 'pawn' or 'c5 square'. These abstractions are only sensitive to key features that define each entity, with invariance to other features. The first and second images are both Caro-Kann despite having almost no pixel similarity in the position of the c-pawn, while the third is Sicilian. **b,** Replay sequences in a spatial task jump abruptly from point to point[116]. Animals ran around a 2D open field (black square), and replay was measured during subsequent rest. Image shows a single replay sequence, collapsing over 340 milliseconds. Color at each pixel is the decoded probability that the replay sequence traversed that point in space.

magnetoencephalography (MEG), which enables the study of tasks with interesting nonspatial relationships that are difficult to explain to rodents[73]. In a nutshell, MEG replay experiments involve three stages. First, participants are presented with experimental stimuli (such as images of objects) in random order, each stimulus repeated many times. The evoked MEG data are used to train decoding models that recognize the spatial pattern of MEG data coding for each stimulus. Second, participants perform a task in which they learn about some relationships between the stimuli (for example, chair always precedes ball). Third, during a time period of interest (such as a resting period), MEG data are analyzed using the trained decoding models to identify spontaneous reactivations of the stimulus representations. Replay is declared when spontaneous reactivations occur in fast sequences which adhere to the relationships learned about in the task.

A number of experiments using this method have found replay in nonspatial tasks. Kurth-Nelson et al[73] designed a directed graph of six nodes, where each node was defined by an object (cat, chair, etc) (Figure 4a). Participants experienced trajectories through this graph, always seeing one object at a time, with the order of objects defined by the graph transitions. The graph itself had no natural spatial embedding, and subjects reported not conceiving of the objects in space, instead using conceptual mnemonics like 'the cat sits on the chair'. After participants had learned about this implicit graph, MEG activity during rest periods spontaneously replayed sequences of object representations, in reverse order of the actual transitions of the graph (Figure 4b). Sequences were time-compressed approximately 20-fold, so an entire sequence lasted about 120 ms. Nonspatial replay also matches several other identifying features of traditional spatial replay, including reward-dependent direction reversal and coincidence with sharp wave ripples[81].

Although these experiments used a simple set of visual objects, hippocampus codes entities as diverse as social position[141], state in abstract MDPs[136,166], time[32], sequence order[1], evidence[101] and auditory pitch[3]. Within hippocampus's switchboard[21,142,144], we could imagine that replay sequences link together entities of different type – even from different levels of abstraction, such as hope and feathers, or equations and planetary orbits.

**First type of composability in replay: attaching objects to role representations**

We described the binding of entities to roles as the first type of composability. If I'm given the set of words {*attacked, alligator, python*} without knowing the role of each item, I don't know who attacked whom. But if I bind each item to its role – {*attacked/action, alligator/patient, python/agent*} – then I understand the meaning.

Human MEG data indicate that replay sequences encode roles bound to objects. Liu et al[81] trained human subjects on a nonspatial task designed to enable decoding of brain representations of roles separately from role-fillers. In the task, subjects learned about a set of eight objects that were arranged into two sequences of four items each. Each object's role could therefore be described by two variables: which sequence it belonged to, and which position (1st, 2nd, 3rd or 4th) it occupied within that sequence (Figure 4c). Subjects performed this task in the MEG scanner, and Liu et al[81] constructed two kinds of neural decoders. The first kind of decoder was trained to recognize each individual object (such as a house). The second was trained to recognize abstract representations of either position or sequence. Abstract position decoders predicted the category label of 1st, 2nd, 3rd or 4th position, given the MEG data evoked by viewing either of the two objects having that position. Likewise, abstract sequence decoders predicted which sequence a viewed object belonged to, invariant to its position. The researchers validated these decoders by testing them with a special cross-validation technique where all instances of a particular object were held out from training. For example, all instances of the 3rd item in the 2nd sequence might be held out from training. In testing, the position decoder successfully predicted that the held-out item was in the 3rd position, and the sequence decoder successfully predicted that the item was in the 2nd sequence.

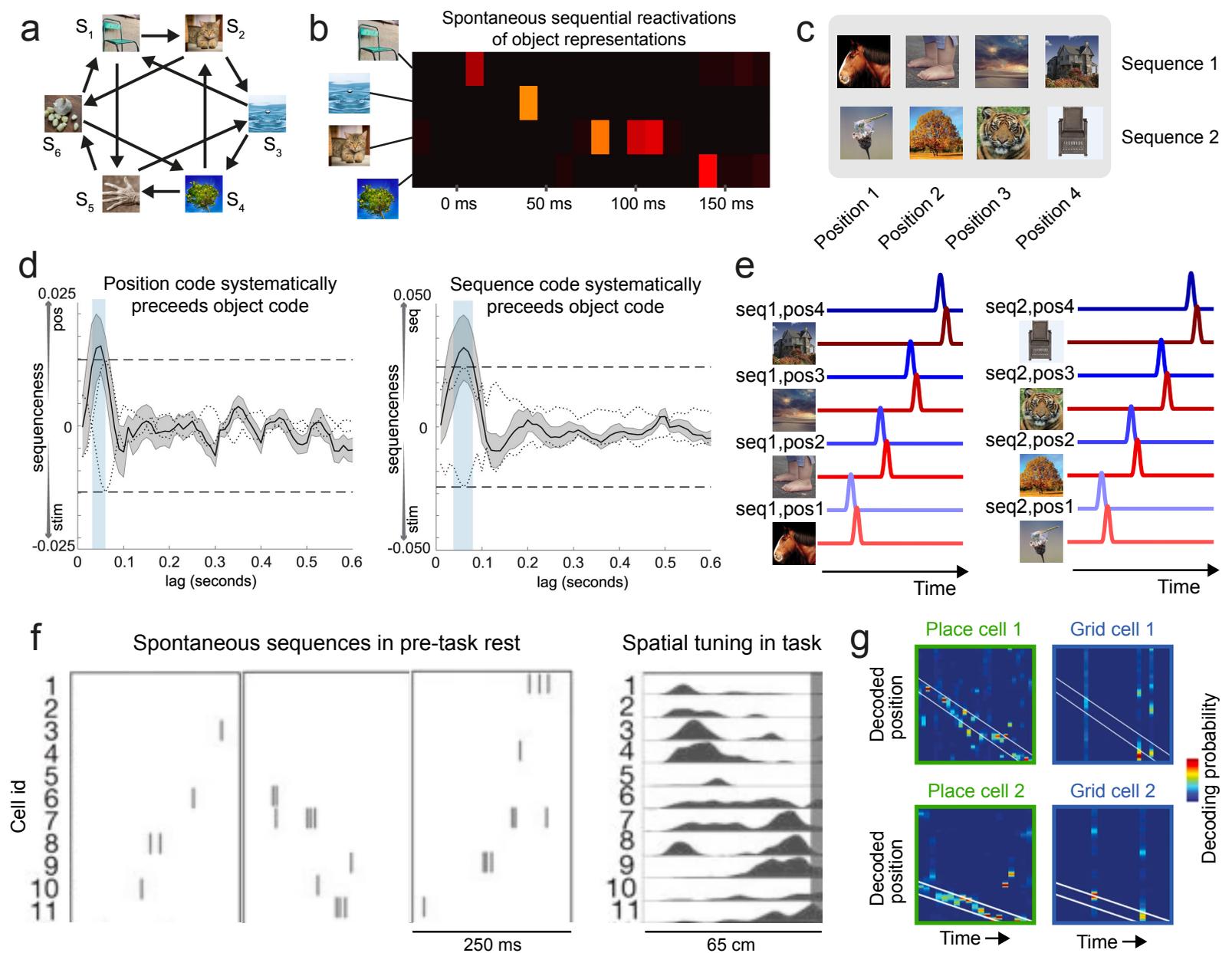

**Figure 4: Nonspatial replay and role representations. a,** In Kurth-Nelson et al[73], subjects learned about the nonspatial relationships between six objects. **b,** Replay sequences were detected with MEG and followed reverse paths in the graph[73]. Color axis shows strength of spontaneous neural reactivations of four stimuli. **c,** In Liu et al[81], subjects learned about a different nonspatial organization: eight objects comprising two sequences. The role of each object in the task could thus be described with two variables: position and sequence. **d,** During a rest period[81], replay spontaneously played out the learned sequences (not shown; similar to panel b). During these replay events, representations of the position and sequence variables also activated, with each role code slightly preceding the replayed representation of the corresponding object. The y-axis is a measure of how consistently the position or sequence code precedes (positive) or follows (negative) the representation of the object. **e,** Schematic of role tags in replay sequences[81]. Colored lines show decoding of spontaneous representations at rest, with y-axis representing strength of decoding. Red lines correspond to concrete objects, which reactivated in fast sequences. Each object was accompanied by representations of the role it played: which sequence it belonged to, and which position it occupied within its sequence (blue lines). **f,** Dragoi and Tonegawa[29] recorded replay sequences that formed trajectories through an environment before the animal's first experience of that environment. Hippocampal cells spontaneously fired in sequences which played coherent trajectories through an as-yet unvisited environment (left; each panel is a replay event recorded before experiencing the new environment). Those sequences were decoded by recording the spatial tunings that the cells subsequently adopted when the animal later experienced the new environment (right). **g,** Replay is coordinated between place cells and grid cells[103]. Two example replay events are shown, with one place cell and one grid cell for each event. Color indicates probability of decoding at each spatial position; white lines are fit to the data.

After constructing these decoding models, MEG activity was analyzed during a rest period that followed task learning. First, using object-specific decoders, spontaneous replay sequences were found to play out representations of the objects within a sequence. These replays coincided with transient increases in power in sharp wave ripple frequency. Second, using the abstract position and sequence decoders, the researchers found that each object in a replay sequence was accompanied (with a characteristic 50 ms lag) by a spontaneous reactivation of that object's abstract sequence and position (Figure 4d,e). In other words, each item in the replay sequence was tagged with its role. The effect is reminiscent of theories like dynamic binding where transient synchronous neural relationships bind roles to contents[35,54,55].

The same role tags that played in alignment with object replay sequences also played out in spontaneous sequences even before the participants experienced the objects. This phenomenon, which the researchers called 'transfer replay', supports the idea that role representations exist separately from role-fillers and are subsequently bound to new role-fillers when they are learned. Transfer replay mirrors 'preplay' in rodents, where spontaneous sequences play out coherent trajectories through an environment before that environment has been experienced[29] (although see [125]) (Figure 4f). The sequences measured in the two experiments were of different kind: Dragoi and Tonegawa[29] measured conjunctive place codes while Liu et al[81] observed role codes. However, neither experiment was set up to detect the other kind of code. It is possible that role representations played out side-by-side with conjunctive representations in Dragoi and Tonegawa[29] but were not measured. Likewise, hypothetical sensory bindings may have played out along with roles in Liu et al[81]. Another unresolved question in both studies is whether the played structure (i.e., linear sequence) was hard-coded by evolution or learned from prior experiences.

The experiments in Liu et al[81] did not afford the anatomical precision to determine where role tags in replay are represented, but grid cells are a prime candidate. As discussed earlier, grid cells code for spatial relationships in spatial tasks and may code for more general relational roles in nonspatial tasks, while hippocampal cells putatively code for the conjunction of role with sensory specifics. In replay, hippocampal cells play out sequences of these conjunctive representations, while grid cells play out corresponding sequences in coordination[103] (Figure 4g). Given that Liu et al[81] found representations of sensory information playing out in coordinated sequences with role-like representations, it is reasonable to think that replay events may include three kinds of representation in coordinated sequences: conjunctive codes in hippocampus (which are place cells in a spatial task), role codes in medial entorhinal cortex and other associated cortical areas, and sensory codes in lateral entorhinal cortex and other associated cortical areas. Future experiments will be needed to test this prediction.

Importantly, the roles investigated in Liu et al[81] were limited to 'which position' and 'which sequence'. The perspective of this paper implies that replay sequences include more general roles like *verb* or *start point*; future experiments will be required to test this prediction. It would be particularly interesting to test for the existence of role tags in replay that could support sophisticated computations. For example, if roles include operations such as *if*, *then* and *else*, then perhaps rudimentary program fragments could be executed in replay sequences. Replay is therefore an intriguing candidate neurophysiological mechanism to implement symbolic computation in the brain[25].

Of course, in the real world, each entity plays different roles in different contexts. A car could be a way of getting to work, a source of scrap metal, or something to stand on. We suspect that replay sequences assemble entities bound to relevant roles according to the context, but understanding the mechanisms will require new experiments.

**Second type of composability in replay: stringing elements into sequences**

In spatial tasks, replay combines fragments of separate spatial experience to play out trajectories that are physically possible but were never experienced[46,115,156]. In nonspatial tasks, the scope for recombination of experience is vastly larger.

Liu et al[81] found that replay composes arbitrary objects into novel sequences using nonspatial relational knowledge. We described this experiment above but omitted one important detail. Participants experienced a set of eight objects: A, B, C, D, A', B', C', D'. (The assignment of actual images, such as a chair, was randomized across participants.) Successful task performance required knowing that the objects were organized into two sequences: <A, B, C, D> and <A', B', C', D'>. But the participants only experienced the objects in a scrambled order, such as <D', B, C', C, D, A', A, B'>. They had previously learned a rule which allowed them to unscramble what they saw into the two true sequences. Behaviorally, subjects performed well at applying the unscrambling rule to derive the true sequences <A, B, C, D> and <A', B', C', D'>. As described in the previous section, subjects had a rest period in the MEG scanner after experiencing the scrambled sequence, and object-specific decoders were used to detect spontaneous reactivations of the eight objects. Surprisingly, participants' brains did not replay sequences in the order of experience. Instead, they played out in the never-seen rule-defined order (Figure 5a). It appears that replay strung a set of items together into a relevant novel compound.

How is this reorganization achieved in the brain? As we saw in the previous section, replay of objects in Liu et al[81] was accompanied by representations of which sequence and which position each object belongs to. Additionally, the position codes replayed spontaneously before the task objects were experienced ('transfer replay'). Taken together, these points of data suggest a potential mechanism for reorganization: the brain has an abstract template for 'items in a sequence'[9,84], and when new items are encountered that fit this pattern, they are attached to the appropriate role in this template, allowing them to play in the correct sequence. However, for a general-purpose compositional system, it would be useful to not only fit new entities into a static template (like 'items in a sequence'), but also to produce new compounds where the roles themselves can be reorganized. For example, here is an English sentence in which not only are the words new, but the syntactic structure – the sentence diagram – is not a copy of any sentence I've previously written.

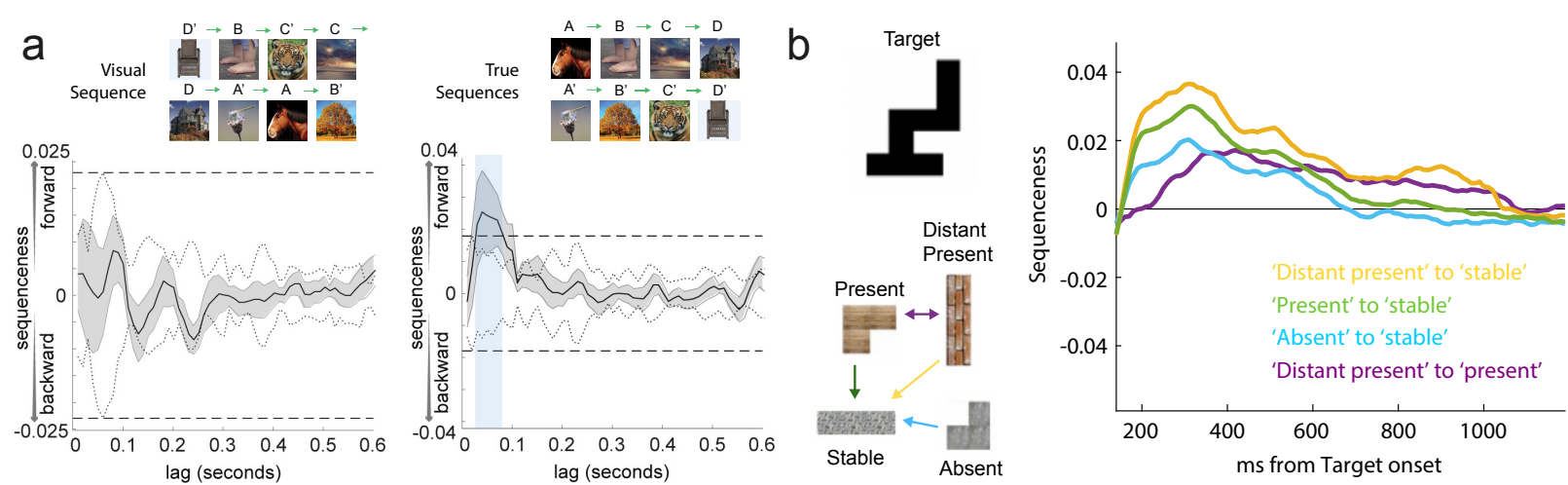

**Figure 5: Replay sequences assemble entities into new compounds. a,** Replay synthesizes novel sequences that are implied by an abstract rule[81]. Human subjects learned a rule that defined how a set of objects should be ordered. When they encountered a new set of objects (the eight objects shown) out-of-order, replay immediately began playing the items in the rule-defined order (True Sequences). There was no replay in the actually-experienced order (Visual Sequence). Each plot shows the strength of forward (positive y) or backward (negative y) sequences at every possible replay speed (x-axis). **b,** In a construction problem, sequences compose building blocks into candidate solutions[123]. In each trial of this task, subjects are presented a black silhouette (Target) and given 3.5 seconds to consider how to build it from four component blocks they previously learned about. The Stable block is part of the solution for every Target (in varying positions). Each other block can be Present or Absent in any given Target. In the MEG signal, spontaneous representations of each block were decoded during the 3.5 second thinking period. The right panel shows the degree of consistent sequences (y-axis) in the spontaneous activations at each moment relative to Target onset (x-axis). Tick labels are the start point of a 1 s sliding window used for sequence analysis. This plot zooms in on the most relevant time epoch. Early in the deliberation period, three kinds of replay sequences appeared (yellow, green and blue traces), each involving the Stable block and one other block. Soon after, sequences emerge that bidirectionally link the two Present blocks (purple trace). Note that sequenceness here measures absolute rather than forward minus backward sequences as plotted in panel a. Negative A->B sequenceness occurs if A is followed by a reduced probability of B. Sequenceness is averaged over the range of element-to-element lags (10-200 ms) typically used for MEG replay analysis.

In Schwartenbeck et al[123], in each trial human participants were shown the silhouette of a complex two dimensional shape (Figure 5b) and given 3.5 seconds to think about how to assemble it from a set of four primitive building blocks they had previously learned about. Each solution involved exactly three blocks. Participants were aware that one particular building block would be present in all silhouettes. We call this block 'Stable'. The block that directly attaches to Stable, in the ground truth solution for a given trial, we call 'Present'. The block attaching to 'Present' we call 'Distant Present'. The fourth block is 'Absent'. To investigate replay signals underlying the mental construction process, the investigators trained classifiers on individual building blocks and measured pairwise sequences between spontaneous reactivations of those representations during the deliberation period, using similar methods as Liu et al[81]. They found significant replay of several kinds of sequences of blocks. For example, detection of Present→Stable replay meant that spontaneous reactivation of Present was consistently followed by Stable. Importantly, forming these replay sequences involved placing blocks into new arrangements, analogous to a new sentence diagram. Another interesting feature of this study is that there was no temporal ordering between blocks, meaning that replay played out sequences constituting a set of elements rather than successive states in an MDP.

Schwartenbeck et al[123] also found that the content of the sequences changed dynamically during the deliberation period. Initially, sequences played out all combinations of Stable with another block. In other words, there were independently sequences of Present→Stable, Distant Present→Stable and Absent→Stable. Shortly after, sequences appeared which combined Distant Present with Present. The researchers argued that each replay sequence can be viewed as sampling a hypothesis about a partial solution of the target configuration, in order to evaluate the hypothesis and gradually resolve uncertainty about the solution. The earliest sequences reflected all combinations of blocks compatible with the prior that the Stable block exists somewhere in the solution. These sequences ostensibly discovered that Present attaches to Stable, after which sequences started to sample the combination of Distant Present with Present, forming the remainder of the solution.

The data from Schwartenbeck et al[123] suggest that not only is replay not constrained to play out the actual order of experience, but it may not even be constrained to play out successive states in an MDP. After viewing each silhouette and thinking about how it could be built from blocks, participants were shown two blocks and asked a yes/no question about whether these blocks would have a particular relationship (such as "Right of") in the solution. They never actually placed blocks to construct a solution. Thus, the task doesn't have an MDP in which the observed replay sequences are successive states. Of course, it is possible that participants construct a fictive internal transition model, and replay consists of rollouts in the fictive model. This seems unlikely given the observed pattern of replay, but would be difficult to fully rule out because nearly any pattern of data could in principle be explained by some fictive model.

One interpretation of the Schwartenbeck data is that sequences traverse paths in a graph of entities where edges are arbitrary relationships between entities. In fact, even constraining replay to traverse coherent paths might not be necessary if each item in a replay sequence is coupled to its role. However, order may provide extra information, for example to de-alias ambiguous or repeated roles; order is apparently also a downstream consequence of the neural mechanisms that generate sequences. More work will be needed to understand the significance of order within sequences.

**Planning with compositional replay**

Reorganization of knowledge in replay sequences may be a mechanism for planning. When animals face a spatial decision, spontaneous neural sequences rapidly sample multiple possible future trajectories[62,66,147], likely as part of a planning and evaluation mechanism[59,105,114,127,148]. Planning in nonspatial tasks is thought to work similarly, sampling compositions of prior knowledge to search for solutions[52,53,57]. We suspect, then, that replay actively constructs new compounds (possibly in the form of program fragments) which represent potential partial solutions to a problem. Sampling compositional hypotheses in replay aligns with the brain's widespread use of sampling for uncertainty representation and inference[11,17,31,44,106,117].

There is of course a continuum between online planning and offline processing of knowledge. When active behavior is required, replay samples sequences relevant to that behavior[115,157]. Without task demands, sampling is less constrained[104,135], presumably to prepare for a range of possible future scenarios. But the organism always has some information about upcoming decisions, so offline processing lies on a spectrum with planning[24,33]. When decisions are not needed imminently, we imagine that replay reviews the knowledge base, deriving new facts and identifying inconsistent beliefs requiring revision, like generalized path integration[96] in conceptual space. When a child learns 'dolphins are mammals'[43,93], offline replay begins to explore and discover the implications, linking and cross-checking it with other knowledge. This takes a while, during which time contradictory beliefs will often be held.

Is replay too noisy to do such intricate computations? Although statistically reliable, the magnitude of sequence effects is often low in absolute terms[73,81]. However, our measurements of replay exaggerate the amount of noise. Not only are our decoders of neural activity far from perfect, but also the brain is doing much besides dealing with the logic of the experimenter's task. Additionally, noise is not all bad; for example, noise drives sampling and exploration[95,106].

**Replay's bidirectional interactions with cortex**

So far we have focused on replay in hippocampus, but hippocampal replay of course functions as part of the larger system of the brain. One very basic point of connection is that the representations in hippocampus derive from cortex. Thus the compositional computations of replay are grounded[48] in cortical representations. (Ongoing replays may in fact help to maintain the tight coordination between hippocampus and cortex[64].)

The contents of replay sequences are also shaped by cortex. For example, when an auditory stimulus is played to rats during sleep, spontaneous reactivations in hippocampus are biased toward places previously associated with that sound[10], with the hippocampal activations immediately preceded by corresponding activations in auditory cortex[119]. Although those experiments measured individual reactivations rather than sequences, replay sequences have also been observed where cortex systematically leads hippocampal representations[60].

Additionally, the contents of replay sequences are influenced by the organism's current goals[102,115,157,161] (Figure 6a). Liu et al[81] found that spontaneous value representations localized to vmPFC were predictive of replay of a rewarded sequence, but not of an unrewarded sequence – following the general principle of frontal areas orchestrating hippocampal memory retrieval[94,98]. We conjecture that, if hippocampal memory is thought of as a probabilistic model of the large joint distribution of experience, cortex may bias retrieval by conditioning that distribution on particular features (for example, reward) to obtain relevant samples. Of course, the degree to which hippocampus itself encodes joint structure between items, versus this information coming from cortex, remains an open question.

In the other direction, hippocampus-to-cortex, our proposal builds on the classic theory of systems consolidation[92,130,158], where hippocampus rapidly stores new experiences and later reactivates them to consolidate the knowledge into a more stable form in cortex. A wealth of evidence indicates a role for replay in consolidation[20,105,131]. Sharp wave ripples, in which replays are nested, drive increased hippocampal-cortical communication and cortical plasticity; while disrupting ripples impairs learning (Figure 6b).

Consolidation theory also posits that replaying different experiences in close proximity allows cortex to learn latent commonalities[58,92,108,110,120]. This semantic knowledge induced in cortex by replay may itself have grammar-like structure[6,7]. The contents of replay are also optimized to cause the learning that will be most relevant for future tasks[2,5,26,82,137].

As described above, newer experiments have identified situations where replay does not strictly recapitulate past experience, but instead generates new trajectories using a learned transition model, which itself is plausibly situated in cortex[46,81,102,115,135,156]. Such model-based replays do not conform to the simplest models of consolidation because they do not faithfully reactivate actual past experiences. They may nevertheless support transfer of knowledge from hippocampus to cortex by training cortex on the statistics of possible trajectories, analogous to Dyna in machine learning[139]. Replaying actual experience can in fact be viewed as a special case of model-based replay using a nonparametric model[150]. There is some evidence that model-based replay has the additional function of contributing to online planning[105].

As model-based replay generalizes consolidation, our proposal further generalizes the idea of model-based replay. We have proposed that not only are replays not limited to playing back actual experience, they are not even limited to playing out trajectories defined by a transition model. Instead, they play out sets of elements along with syntactic roles that define how the elements function together as a whole. Furthermore, instead of directly updating a value function or policy (as in Dyna), we suggest that replay discovers new knowledge of a more general form that does not necessarily have any immediate impact on the policy. An interesting implication is that consolidation mechanisms cause cortex to learn about the new knowledge discovered by composing elements into sequences.

The bidirectional interaction between replay and cortex sets up the possibility for a positive feedback loop[77,110]. According to our theory, replay composes existing entities into sequences, and new knowledge is derived from the meaning of the sequence as a whole. If that new knowledge is baked into cortex[145], it could subsequently function as a single element in higher-order replay sequences (i.e., sequences composed of pieces of knowledge which were previously derived from other sequences). The loop might produce a hierarchy of increasingly elaborate concepts. A similar feedback loop could operate at a faster timescale, too. Some neurons in prefrontal cortex are tuned to specific replay sequences[12], and if these neurons feed back into hippocampus as entities, then higher-order sequences could be composed within hundreds of milliseconds.

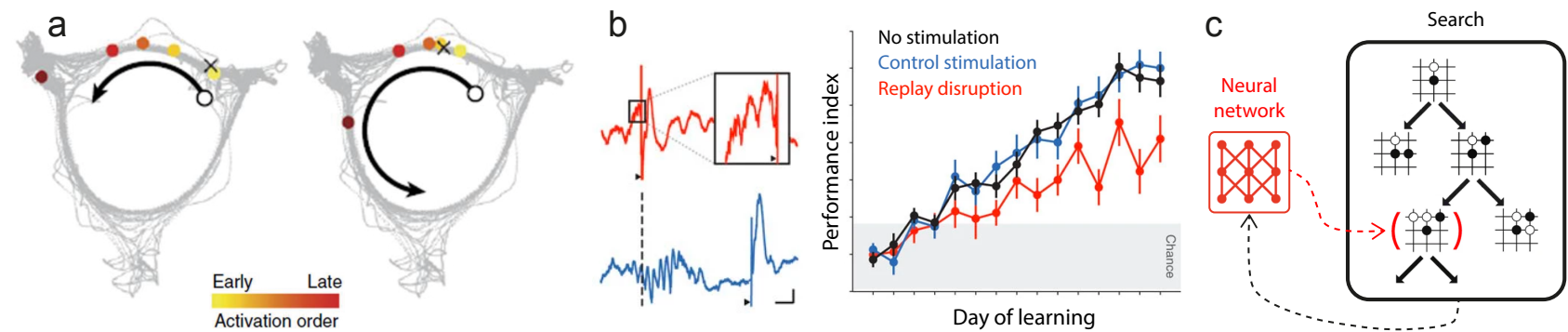

**Figure 6: Hybrid systems. a,** Current goals bias replay sequences. In Wikenheiser and Redish[157], rats ran in a circle, sometimes electing to pause at one of three equally-spaced feeders (upper-left, upper-right, and bottom). Replay sequences (colored dots) traversed paths ahead of the animal. In trials when the rat would subsequently stop at a particular feeder, the prospective sequences also terminated at that feeder (left panel). When the rat did not plan to stop, sequences skipped the feeder (right panel). 'X' is the animal's location. **b,** Blocking replay impairs learning. Replay sequences are nested within sharp wave ripples, the high frequency oscillation shown in blue trace and inset of red trace. Girardeau et al[39] performed online detection of ripples during a rest period after each day's learning session, and injected current to disrupt replay events at their initiation (upper left panel). In the control condition (blue), current was injected 100 ms after the detected ripple (lower left panel). The calibration bars show 20 ms on the x-axis and 0.2 mV on the y-axis. Disrupting replay during rest impaired acquisition of the task (right panel; red disrupted learning curve below blue control and black no current injection). **c,** Simplified diagram of AlphaGo[126], which has a neural network component (red) and a discrete search component (black). The network predicts which moves are good in each board position. This constrains the search mechanism, reducing the combinatorial explosion of the game. The results of searches are ultimately used to train the neural network, creating a feedback loop.

**Experimental predictions**

Our perspective is that replay sequences sample structured compositions of existing elements to derive new knowledge. The most obvious experimental implication is that when replay is disrupted, the derivation of new knowledge through composition should be impaired. This could be tested using a task where subjects are taught about novel elements and how those elements work together. After teaching subjects about these elements, the experiment would require detecting the initiation of replay and immediately disrupting it[39,59], with the prediction that disrupted individuals would subsequently demonstrate less knowledge about implications of the rules. Such an experiment is challenging because reasoning in rodents is limited, and disrupting replay in humans has never been attempted. Experiments in rodents would require design of tasks to access compositional cognition, perhaps by training on a small set of conditional rules. Disruption in humans might be possible with ultrasound or TMS, especially if replay initiation could be predicted slightly in advance using machine learning. In disruption experiments, careful controls for nonspecific effects such as impaired memory for task-related information would be essential.

Disruption experiments will be vital for establishing whether replay has a causal function in compositional computation. Even when replay appears sophisticated, it is possible that this complexity is epiphenomenal, a kind of 'exhaust fume' for computations actually performed elsewhere. However, there are reasons to be optimistic that compositional replay is in fact causal. First, as discussed in the previous section, disruption experiments in rodents have shown that replay is causally involved in learning and decision-making. Second, replay has pieces of the machinery – like the role codes in Liu et al[81] – that would be useful for performing compositional computation. Third, replay relevant to a novel problem appears within hundreds of milliseconds of the onset of that problem, as in Schwartenbeck et al[123].

Aside from disruption experiments, another prediction is that after knowledge is derived by replay, it is subsequently available for use in other neural computations. For example, novice players could be tasked with solving a chess puzzle where the king is in check by a bishop (of course a real experiment would be designed to eliminate the confounds inherent to chess). MEG decoding models would be trained to recognize representations of the pieces as well as their relational roles. As subjects consider the puzzle, we predict that replay sequences explore various organized combinations of the pieces, examining the implications of their interactions. At some point during deliberation, a sequence is constructed that contains the king and bishop as elements and the bishop's moving pattern as a relation. This sequence implies that the king is in check. After this sequence appears (but not before), the brain has knowledge of the king being in check, and this knowledge should manifest in several ways. It might be possible to decode the representation of 'king in check' from the brain. Subsequent replay sequences might be constrained to evaluate only moves that escape check. And if forced to move, subjects might behaviorally reflect the knowledge of the king being in check. Although such a correlation would not prove a causal role for replay, it would be suggestive if the temporal relationship were consistent, with knowledge of 'king in check' appearing at a short predictable delay after replay sequences containing the king and bishop.

Finally, we have hypothesized that representations of items in a replay sequence are accompanied by representations of relational roles that describe how the items fit together to produce an overall meaning. As described earlier, Liu et al[81] found a special case of these relational roles in replay: representations of 'which sequence does this item belong to' and 'which position does this item occupy'. Future experiments can test more general roles. For example, if participants are asked to name the relationship between two distant relatives in a family tree, do replay sequences sample possible ways of linking them (like 'mother's father's brother'), and are replayed representations of individuals accompanied by representations of familial roles like 'brother' or 'stepdaughter'? Again, a real experiment should use an appropriate grammar of relationships to reduce confounds.

## Replay in the brain and AI

In this final section, we consider the implications of our proposal for the relationship between hippocampal replay and machine learning. Replay has been a key point of connection between AI and neuroscience[51,78,97,159]. Replay in the brain is commonly compared to the technique of 'experience replay', where an agent stores its observations and later retrieves them for offline learning. This idea fits well with neuroscientific results where replay appeared to recapitulate actual experience. It can also accommodate the observations, discussed in earlier sections, that replay generates novel sequences. A space of techniques in machine learning use experience to train a forward model of the world's dynamics, and then sample trajectories from the model to train a reactive policy[124,139].

But in this paper we have proposed that replay in the brain has a more general function: deriving new knowledge through compositional computation. So how should we view the correspondence to AI? In this section we suggest that replay, as a kind of compositional computation embedded in the larger brain system including cortex (which behaves in many ways like a deep neural network[68,87,138,162]), should be mapped to machine learning techniques that hybridize compositional computation with deep learning.

While humans learn new concepts or tasks from just a few examples, standard deep learning architectures cannot do this because they do not generalize effectively from their past experience[75]. At the same time, explicitly compositional AI systems without deep networks reason about situations far outside their experience[75,100,146] but typically lack the abilities of neural networks to learn from raw data and scale to large problem sizes. Due to this complementarity, there have been many efforts to hybridize deep learning with compositional ingredients[8,28,34,88,126].

We outline a possible mapping between such hybrid machine learning systems and replay, following five principles:
(1)     Neural networks are used to prune large compositional search spaces.
(2)     New knowledge is discovered by search.

(3)    A positive feedback loop results from networks informing search and search improving networks.
(4)    Compositional operations live along a spectrum from hard-coded to emergent.
(5)    Entities are grounded in neural network representations.

Consider the example of AlphaGo[126] (Figure 6c). Part of AlphaGo is a handcrafted algorithm that searches through trees of possible future moves in order to find move sequences most likely to lead to a win. Another part of the system is a neural network that predicts which moves are likely to be good. The search algorithm prioritizes moves that are preferred by the network, reducing combinatorial explosion and drastically improving search efficiency.

AlphaGo's search is a special case of compositional computing, because it is made up of discrete elements linked in organized ways to derive new knowledge. Viewing AlphaGo through this lens suggests a hypothesis for neuroscience: that cortex predicts which combinatorial arrangements in replay sequences are useful and biases hippocampus toward generating these sequences. Although the syntax of AlphaGo's search only deals with board positions linked by moves, neurally-guided search is also effective in many other machine learning contexts, such as searching through the space of programs expressed in a programming language[34,65,112].

AlphaGo uses the results of search both to make immediate decisions and to train its neural network. When playing a game, AlphaGo selects each move according to a search. But to train the neural network, AlphaGo plays millions of games against itself, and the network is trained to prefer moves that lead to winning outcomes. In this way, new knowledge discovered by the search process is gradually baked in to the network. As AlphaGo's network improves, the quality of searches rises, which further improves the network in a positive feedback loop. While the effect of search to improve AlphaGo's network is indirect (via self-play), other related AI systems[34,47] train a neural network directly on knowledge discovered by search.

Across the large family of methods that hybridize neural networks with compositional representation, there is a wide range of design choices about the degree to which the compositional component is handcrafted versus learned implicitly. At one extreme, AlphaGo uses a fully handcrafted search algorithm and searches through the space of idealized board positions. At the other extreme, training unstructured neural networks on certain distributions of data leads to the emergence of structured compositional operations within the dynamics of the network[45,109,153]. Between these endpoints there are many compromises.

MuZero[122] has a similar architecture to AlphaGo but is not endowed with knowledge of any game's rules. Instead, it learns from experience to transform its inputs into useful representations and to predict the environment's dynamics. Thus, like replay but unlike AlphaGo, MuZero grounds the elements that are recombined through search. However, unlike replay, neither AlphaGo nor MuZero have a rich syntax. Elements are combined only into chains of moves that follow one another in gameplay. An interesting direction in AI research is combining MuZero-style grounded entities and roles with syntactically rich composition.

Neural networks with attention[4,42,151] form another midpoint. Attention mixes together multiple inputs according to a set of weights, such that the inputs with low weight are filtered out and those with high weight are retained. The weights are calculated on-the-fly as a learned function of data. Neural networks with attention have set new standards for performance in many areas of machine learning[15,63,152]. An important observation is that attention gives rise to role-filler binding[14,107,121,129,155]. This is because a role can induce weights that attend selectively to the filler, thereby accessing its contents. If the contents of the filler change, the weights attend to that new data. Or, to rephrase in the language we've been using for replay, attention grants the first type of composability. In the context of replay, it's curious that attention-based neural networks, despite their amazing successes, often fail at structured reasoning[74,89], which depends on composing elements into compounds. Another exciting project in AI research is adding replay-like composition mechanisms to attention-based neural networks.

## Conclusion

The world is too complex for an agent to force all its beliefs to be immediately consistent with one another – we still can't reconcile gravity and quantum mechanics, for example, while having them as separate theories is highly useful. Real intelligence therefore requires understanding things in multiple ways, with a variety of metaphors or views on a problem that are kept separate from one another. But these different conceptualizations need to be sometimes put into contact in order to find new symmetries and deeper understanding. In many contexts, interesting structure arises from alternation between these two modes of a system, one where elements remain relatively separated and another where interactions are frequent[70,132,164]. Perhaps the hybridization of replay with cortex serves this purpose. The compositional system of replay maintains separate pieces of knowledge while cortex extracts semantic abstractions. In this way replay may contribute to the brain's capacity for open-ended learning and creativity[134].

**Acknowledgements.** We thank Zach Duer, Marc Guitart-Masip, Jess Hamrick, Kim Stachenfeld and Steve Sullivan for insightful discussions and comments on a draft of the manuscript. ZKN was funded by DeepMind Technologies.